\documentclass[12pt]{article}

\usepackage{graphicx}
\usepackage{float}
\usepackage{cite}

\def\gtwid{\mathrel{\raise.3ex\hbox{$>$\kern-.75em\lower1ex\hbox{$\sim$}}}}
\def\ltwid{\mathrel{\raise.3ex\hbox{$<$\kern-.75em\lower1ex\hbox{$\sim$}}}}
\def\square{\kern1pt\vbox{\hrule height 1.2pt\hbox{\vrule width 1.2pt\hskip 3pt
   \vbox{\vskip 6pt}\hskip 3pt\vrule width 0.6pt}\hrule height 0.6pt}\kern1pt}

\begin{document}

\begin{titlepage}

\begin{flushright}
UFIFT-QG-18-03
\end{flushright}

\vskip 2cm

\begin{center}
{\bf The Case for Nonlocal Modifications of Gravity}
\end{center}

\vskip 1cm

\begin{center}
R. P. Woodard$^{\dagger}$
\end{center}

\vskip .5cm

\begin{center}
\it{Department of Physics, University of Florida,\\
Gainesville, FL 32611, UNITED STATES}
\end{center}

\vspace{1cm}

\begin{center}
ABSTRACT
\end{center}
The huge amounts of undetected and exotic dark matter and dark 
energy needed to make general relativity work on large scales
argue that we should investigate modifications of gravity. The
only stable, metric-based and invariant alternative to general
relativity is $f(R)$ models. These models can explain primordial
inflation, but they cannot dispense with either dark matter
or dark energy. I advocate nonlocal modifications of gravity,
not as new fundamental theories but rather as the gravitational 
vacuum polarization engendered by infrared quanta produced 
during primordial inflation. I also discuss some of the many 
objections which have been raised to this idea. 

\begin{flushleft}
\end{flushleft}

\begin{flushleft}
$^{\dagger}$ e-mail: woodard@phys.ufl.edu
\end{flushleft}

\end{titlepage}

\section{Introduction}

The case for nonlocal modifications of gravity is easy to make:
\begin{enumerate}
\item{The amount of ``dark'' stress-energy needed to make general
relativity work strains credulity.}
\item{The only metric-based, local, generally coordinate invariant
and potentially stable alternative to general relativity is $f(R)$
models \cite{Woodard:2006nt}. These can explain primordial inflation
\cite{Starobinsky:1980te}, but neither cosmic structures nor the
current phase of cosmic acceleration.}
\item{Although fundamental nonlocality seems problematic, nonlocal 
corrections to the effective field equations from loops of massless
particles can give macroscopic effects, and those associated with 
the vast amount of inflationary particle production become
nonperturbatively strong.}
\end{enumerate}
I discuss each of these points in sections 2, 3 and 4, respectively. 
Section 5 reviews four of the criticisms which have been raised to
my work. My conclusions comprise section 6. 

\section{Shortcomings of Dark Matter and Energy}

Einstein's equation,
\begin{equation}
G_{\mu\nu} = 8\pi G T_{\mu\nu} \; , 
\end{equation}
relates second and lower derivatives of the metric to the stress-energy
tensor $T_{\mu\nu}$. No matter what metric $g_{\mu\nu}$ you want there is 
a $T_{\mu\nu}$ which makes the equation true; general relativity is only 
tested when both sides are known. We do know both sides of the equation for 
stars, but that ceases to be true on larger scales. Dark matter is needed 
to explain galaxies and galaxy clusters, while dark energy is invoked to 
explain the current phase of cosmic acceleration. Dark energy at a scale 
$10^{55}$ higher is the usual explanation for the early phase of 
accelerated expansion known as primordial inflation. In this section I 
review why these explanations are problematic.

\subsection{The willing suspension of disbelief}

Let me start with the sheer magnitude of exotic material which is required 
right now. We are told that only 4.6\% of the current energy density of the 
universe can consist of anything we have ever detected in a laboratory. In 
order to explain cosmic structures general relativity needs approximately 23\% 
of the current energy density to consist of ``dark matter'' which has the 
equation of state of nonrelativistic matter but only interacts weakly. 
Explaining the current phase of cosmic acceleration requires that a whopping 
72\% of the current energy must consist of ``dark energy'' which has the 
equation of state of vacuum energy and interacts at most weakly.

None of this stuff has been seen, except gravitationally. The properties of
dark energy do not require that it be visible in Earth-bound labs but dark 
matter should be, and it has not shown up. Two WIMP searches of unprecedented 
sensitivity reported last year: neither the Xenon 1-ton experiment 
\cite{Aprile:2017iyp} nor PandaX-II \cite{Cui:2017nnn} has detected anything. 
The continued failure to find dark matter has shaken the faith of even some 
passionate believers. To be sure, certain candidates such as Axions are still 
viable \cite{Du:2018uak}. There is also very interesting recent work 
\cite{Garcia-Bellido:2017imq} on the old idea that dark might might not 
be exotic at all, but consists instead of normal matter which formed 
primordial black holes during inflation \cite{Dolgov:1992pu}.

\subsection{Unexplained regularities of cosmic structures}

For me, the real problem with dark matter is its failure to explain observed
regularities in cosmic structures. These {\it are} well explained by Milgrom's
MOdified Newtonian Dynamics, or MOND \cite{Milgrom:1983ca,Milgrom:1983pn,
Milgrom:1983zz} which can be viewed as the static, weak field limit of a 
modified gravity theory \cite{Sanders:2002pf,Famaey:2011kh}. Among the 
regularities it explains are \cite{Sanders:2008iy}:
\begin{itemize}
\item{{\bf The Baryonic Tully-Fisher Relation} $v^4 = a_0 G M$ between the
asymptotic rotational velocity $v$ and the baryonic mass $M$ of some structure,
where $a_0 \simeq 1.2 \times 10^{-10}~{\rm m/s}^2$ \cite{McGaugh:2005qe};}
\item{{\bf Milgrom's Law} that dark matter always starts being necessary when
the acceleration drops below $a_0$ \cite{Kaplinghat:2001me};}
\item{{\bf Freeman's Law} $G \Sigma < a_0$ for the surface density $\Sigma$
\cite{Freeman:1970mx}; and}
\item{{\bf Sancisi's Law} that features in luminous matter follow features in
rotation curves and vice versa \cite{Sancisi:2003xt}.}
\end{itemize}
Especially impressive is the recent work by McGaugh and collaborators on the
universal relation which seems to exist between the observed radial acceleration
and that predicted using only baryons \cite{McGaugh:2016leg,Lelli:2017vgz}. This 
does not accord well with the idea that dark matter is five times more prevalent 
than baryonic matter. With general relativity plus dark matter one has to wonder, 
why is the baryonic matter tail wagging the dark matter dog?

The bottom line is that MOND provides too good a fit to evolved structures
to be an accident. Either general relativity with dark matter approaches 
MOND as some kind of hitherto unrecognized attractor solution or else there 
is no dark matter and MOND represents the nonrelativistic, static limit of 
some modified gravity theory. Either possibility is fascinating, and I don't 
think anyone can honestly claim to know which is correct right now. Because 
I work in gravity I have chosen to explore the second possibility, 

\subsection{Fine tuning problems}

I think fine tuning is the worst problem for dark energy, and for primordial 
inflation. The usual explanation for both things is general relativity plus a
minimally coupled scalar whose potential drives acceleration,
\begin{equation}
\mathcal{L} = \frac{R \sqrt{-g}}{16\pi G} - \frac12 \partial_{\mu} \varphi
\partial_{\nu} \varphi g^{\mu\nu} \sqrt{-g} - V(\varphi) \sqrt{-g} \; . 
\label{GR+MMCS}
\end{equation}
There is no question that this sort of model can support the required 
expansion histories because there is a closed form procedure for 
constructing the potential \cite{Tsamis:1997rk,Saini:1999ba,Nojiri:2005pu,
Woodard:2006nt,Guo:2006ab}. Suppose the desired geometry takes the form, 
\begin{equation}
ds^2 = -dt^2 + a^2(t) d\vec{x} \!\cdot\! d\vec{x} \quad \Longrightarrow 
\quad H(t) \equiv \frac{\dot{a}}{a} \;\; , \;\; \epsilon(t) \equiv -
\frac{\dot{H}}{H^2} \; . \label{FLRW}
\end{equation}
We assume the scalar depends only on time $\varphi = \varphi_0(t)$. The two
nontrivial Einstein equations are,
\begin{eqnarray} 3 H^2 & = & 8\pi G \Bigl[ \frac12 \dot{\varphi}_0^2 + 
V(\varphi_0)\Bigr] \; , \label{Ein1} \\
-2 \dot{H} - 3 H^2 & = & 8\pi G \Bigl[ \frac12 \dot{\varphi}_0^2 - V(\varphi_0)
\Bigr] \; . \label{Ein2}
\end{eqnarray}
Adding (\ref{Ein1}) and (\ref{Ein2}) gives an equation we can solve for the
scalar, up to an initial condition and a sign choice,
\begin{equation}
-2 \dot{H} = 8\pi G \dot{\varphi}_0^2 \quad \Longrightarrow \quad \varphi_0(t) 
= \varphi_0(t_i) \pm \int_{t_i}^{t} \!\!\! ds \, 
\sqrt{\frac{-2 \dot{H}(s)}{8\pi G}} \; . \label{scalar}
\end{equation}
Assuming $\dot{H}(t)$ is negative-definite, we can invert this (at least 
numerically) to determine the time as a function of the scalar $\tau(\varphi)$.
Now subtract (\ref{Ein2}) from (\ref{Ein1}) and solve for the potential,
\begin{equation}
V(\varphi) = \frac{\dot{H}(t) \!+\! 3 H^2(t)}{8\pi G} 
\Biggr\vert_{t = \tau(\varphi)} \; . \label{potential}
\end{equation}

The construction (\ref{FLRW}-\ref{potential}) leaves no doubt that scalar
potential models (\ref{GR+MMCS}) can support any expansion history with
$\dot{H}(t) < 0$, but we are left wondering {\it who ordered that?} More
quantitative questions abound:
\begin{itemize}
\item{Why is $\varphi(t,\vec{x}) \sim \varphi_0(t)$ so spatially
homogeneous?}
\item{Why is $G^2 V(\varphi_0) \sim 10^{-122}$ so small?}
\item{Why is no 5th force observed?}
\end{itemize}
For primordial inflation the degree of fine tuning needed to get inflation
to start, and the tendency to lose predictivity \cite{Ijjas:2013vea} has
led to considerable angst within the community \cite{Guth:2013sya,
Linde:2014nna,Ijjas:2014nta}. There is an additional problem associated with
the need to couple the inflation $\varphi$ to ordinary matter to make
reheating efficient. On de Sitter background ($\epsilon = 0$) the resulting 
cosmological Coleman-Weinberg potentials turn out to depend in a complicated 
way on the dimensionless ratio of $\varphi/H$. These potentials are not
Planck-suppressed and they cannot be fine-tuned away because the factors of
``$H$'' are not even local for a general metric \cite{Miao:2015oba}. This
seems to have a disastrous effect on inflation \cite{Liao:2018sci}. 

I should mention that there are two reasonable alternatives to (\ref{GR+MMCS})
for primordial inflation which avoid some of the fine-tuning problems. One of 
these models employs the Higgs as the inflaton, but with a huge conformal 
coupling \cite{Bezrukov:2007ep}. The other is a modified gravity theory based 
on adding a large $R^2$ term to the Hilbert action \cite{Starobinsky:1980te}.

\section{Options for Modifying Gravity}

Modified gravity theories can be classified based on the answers to three
questions:
\begin{enumerate}
\item{Is the gravitational force entirely carried by the metric or are 
other fields involved?}
\item{Is full general coordinate invariance preserved?}
\item{Are the field equations local or nonlocal?}
\end{enumerate}
In this section I will discuss metric-based modifications of gravity 
which preserve full general coordinate invariance. An important theorem 
restricts local, stable theories of this type to just $f(R)$ models
\cite{Woodard:2006nt}. I begin by explaining why $f(R)$ models cannot
replace either dark energy or dark matter. I then review the problems 
associated with fundamental nonlocality.

\subsection{Problems with $f(R)$ models}

I have already mentioned that an $f(R)$ model can give primordial 
inflation \cite{Starobinsky:1980te}. The same is not true for explaining
the current phase of late time acceleration. The data tell us that the
$\Lambda$CDM expansion history seems to be correct \cite{Abbott:2017wau,
Troxel:2017xyo}, however, the only stable $f(R)$ model which reproduces 
the $\Lambda$CDM expansion history is $f(R) = R - 2 \Lambda$ 
\cite{Dunsby:2010wg}. This means that any $f(R)$ model which replaces 
dark energy is bound to show discrepancies with the data at 0th order, 
without even worrying about perturbations.

To see the problem replacing dark matter, consider the geometry of a 
static, spherically symmetric and nearly flat geometry,
\begin{equation}
ds^2 = -\Bigl[1 + b(r)\Bigr] dt^2 + \frac{dr^2}{1 \!+\! a(r)} + 
r^2 d\Omega^2 \; . \label{static}
\end{equation}
Suppose $M(r)$ represents the mass enclosed within radius $r$. If this
system is a low surface brightness galaxy within the MOND regime for all 
$r$, then the baryonic Tully-Fisher relation says the potential $b(r)$ 
must obey \cite{Deffayet:2011sk},
\begin{equation}
v^4(r) = a_0 G M(r) = \Bigl[ \frac12 r b'(r)\Bigr]^2 \; . \label{Tully}
\end{equation}
Because $M(r)$ is an integral over the mass density $\rho(r)$, we can
recover what must be the weak field, static limit of the MOND equation
for $b(r)$,
\begin{equation}
\frac{\delta S}{\delta b(r)} = \frac1{32 \pi a_0 G} \, \partial_r 
\Bigl[ r b'(r) \Bigr]^2 - \frac12 r^2 \rho(r) = 0 \; . \label{beqn}
\end{equation}
That equation (\ref{beqn}) cannot have come from any $f(R)$ model is 
obvious from the weak field expansion of the Ricci scalar,
\begin{equation}
R = -b'' + \frac{2 (a' \!-\! b')}{r} + \frac{2 a}{r^2} \; . \label{Ricci}
\end{equation}
Note that the problem is fundamental, and has nothing to do with the
weak field expansion. Ricci scalars have two derivatives, or factors of $1/r$,
whereas the desired field equation (\ref{beqn}) has three derivatives. 

\subsection{Problems with fundamental nonlocality}

I think it would be fair to say that Sir Isaac Newton disapproved of
nonlocal equations of motion. He denounced it as so \cite{SIN},
\begin{quote}
{\it great an Absurdity that I believe no Man who has in philosophical
Matters a competent Faculty of thinking can ever fall into it.}
\end{quote}
Now I know that some people at this conference, who {\it do} have a
competent faculty of thinking, support fundamental nonlocality. Without
engaging in Newton's vituperation, let me explain why I share the great
man's doubts about the subject.

Ostrogradsky's theorem states that nondegenerate higher derivative 
models have extra degrees of freedom, essentially half of which carry
negative kinetic energy \cite{Woodard:2015zca}. When these sorts of
theories have interactions among continuum fields they develop a crazy 
time dependence in which the positive energy degrees of freedom become
infinitely excited by infinitely exciting the negative energy ones. 
Some nonlocal theories can avoid this, but not the type favored by people
here, which is based on entire functions of the derivative operator. An 
entire function is {\it defined} to converge to its Taylor series 
expansion, so we know one can view the theories of interest as the limits 
of sequences of ever-higher derivative models \cite{Eliezer:1989cr}. The 
theories in such a sequence become {\it more} unstable, not less, as the 
number of higher derivatives increases. In the fully nonlocal limit one 
can specify the dynamical variable {\it arbitrarily} within any finite 
coordinate range, adjusting the variable outside this range to make the 
equation true. Assertions to the contrary are often based on working 
perturbatively in Euclidean momentum space. This amounts to assuming
away the problem because the wild time dependence precludes the 
existence of temporal Fourier transforms in the first place. 

People who claim to have solved the problem of extra, and unstable, 
initial value data sometimes ask me for an example of a system on 
which they might apply it. As it happens, I spent the better part of 
a year trying to come up with a good solution to the problem of
using the scalar power spectrum of primordial inflation to reconstruct 
the first slow roll parameter $\epsilon$, regarded as a function of
the number of e-foldings $n$, for the case where there are features. A 
simplified version of this problem takes the form of a linear 
integro-differential equation for $\ln[\epsilon(n)]$ 
\cite{Brooker:2017kjd,Brooker:2017kij},
\begin{equation}
\Bigl[1 + G(1) \partial_n \Bigr] \ln[\epsilon(n)] - \int_{0}^{n} \!\!\!
dm \, \Bigl[ \partial_m^2 \!+\! 3 \partial_m \Bigr] \ln[\epsilon(m)]
\!\times\! G(e^{m-n}) = f(n) \; , \label{example}
\end{equation}
where the function $G(x)$ is,
\begin{equation}
G(x) \equiv \frac12 (x \!+\! x^3) \sin\Bigl[ \frac{2}{x} \!-\! 2 
{\rm arctan}\Bigl(\frac1{x}\Bigr) \Bigr] \; . \label{Gdef}
\end{equation}
We never did get a really satisfactory solution, precisely because 
of the Ostrogradskian instability. Those of you who think this is 
no issue, please solve my problem and then we can talk.

Before closing I should mention the claims of another faction of those
who believe in fundamental nonlocality, and also higher derivative 
theories. These people acknowledge the classical problem but assert 
that it can be evaded by clever alternate quantizations. The details 
don't matter much because all such claims suffer from the same problem 
of giving up the classical Correspondence Limit. Of course that must 
be the case because the classical theory has negative energy field
configurations whereas the spectrum of the alternate quantization does
not. Physics is ultimately an experimental subject and if someone 
advanced this idea for anything other than gravity I would agree to
let experiment decide the issue. But the {\it only} low energy 
gravitational data we have, or ever will have, is from classical
general relativity. It is a bad bargain to throw that away in the 
search for something you call ``quantum gravity.'' 

\section{Modified Gravity as Vacuum Polarization}

In criticizing fundamental nonlocality it might be thought that I have
undercut the case I wish to make. However, there is a completely acceptable
type of nonlocality in the form of quantum corrections to the effective 
field equations. I first discuss how loops of massless particles can give
macroscopic effects, even in flat space. Then I discuss why primordial
inflation might produce even stronger effects. This is followed by a 
review of corrections to electromagnetism and to gravitation which become
nonperturbatively strong during a prolonged phase of primordial inflation.
The section closes by reviewing the proposal that such effects might 
perhaps provide a model for primordial inflation.

\subsection{Macroscopic nonlocality in flat space QED}

We all know how electron-positron loops cause electrodynamic forces to 
become stronger at short distances. Static corrections to the Coulomb 
potential $\Phi(r)$ of an electron are described by the nonlocal equation,
\begin{equation}
-\nabla^2 \Biggl[ \Phi(r) + \frac1{2 \pi^2 r} \! \int_0^{\infty} \!\!\! dk
\, k \sin(k r) \chi_e(k) \widetilde{\Phi}(k) \Biggr]
= -e \delta^3(\vec{r}) \; . \label{MaxQuant}
\end{equation}
Here $\widetilde{\Phi}(k)$ is the spatial Fourier transform of $\Phi(r)$
and $\chi_e(k)$ is the one loop contribution to the electric susceptibility,
\begin{equation}
\chi_e(k) = \delta \chi_e + \frac{4 \alpha}{\pi} \! \int_{0}^{1} \!\!\!
dx \, x (1 \!-\! x) \Biggl\{ \ln(2 \Lambda) \!-\! 1 \!-\! \frac12 
\ln\Bigl[ m_e^2 + x (1 \!-\! x) k^2 \Bigr] \Biggr\} \; , \label{Esusc}
\end{equation}
where $\Lambda$ is the ultraviolet cutoff and $\delta \chi_e$ is the 
counterterm. As long as the electron mass $m_e$ is nonzero we would choose
$\delta \chi_e$ to make the susceptibility vanish at $k = 0$,
\begin{equation}
\chi_e(0) = 0 \quad \Longrightarrow \quad \chi_e(k) = -\frac{2\alpha}{\pi} 
\! \int_{0}^{1} \!\!\! dx \, x (1 \!-\! x) \ln\Biggl[1 + x (1 \!-\! x) 
\frac{k^2}{m_e^2} \Biggr] \; . \label{normalEsusc}
\end{equation}
Of course this means that there is no correction to the classical result
for large $r$, however, the small $r$ potential experiences a logarithmic 
enhancement,
\begin{equation}
r \ltwid \frac1{m_e} \quad \Longrightarrow \quad \Phi(r) = -\frac{e}{4\pi r}
\Biggl[1 + \frac{2 \alpha}{3 \pi} \ln\Bigl(\frac1{m_e r}\Bigr) + \dots
\Biggr] \; . \label{UVrunning}
\end{equation}

Now suppose the electron mass vanished. In this case we could no longer 
choose $\delta \chi_e$ to make the susceptibility vanish at $k = 0$. We
would instead choose some renormalization length scale $R$,
\begin{equation}
m_e = 0 \quad \Longrightarrow \quad \chi_e(k) = -\frac{2 \alpha}{\pi} \!
\int_{0}^{1} \!\!\! dx \, x (1 \!-\! x) \Bigl[x (1 \!-\! x) k^2 R^2\Bigr]
\; . \label{masslessEsusc}
\end{equation}
In this case the potential shows logarithmic corrections for all $r$,
\begin{equation}
\forall r \quad \Longrightarrow \quad \Phi(r) = -\frac{e}{4 \pi r} 
\Biggl[1 + \frac{2 \alpha}{3\pi} \ln\Bigl(\frac{R}{r}\Bigr) + \dots 
\Biggr] \; . \label{IRrunning}
\end{equation}
At small $r$ expression (\ref{IRrunning}) exhibits the same enhancement as
for a massive electron (\ref{UVrunning}). However, for {\it large} radius
the potential is weakened, and the effect becomes nonperturbatively strong
for $r \gg R$. Because the effective coupling is weakened for large $r$,
we can sum the sequence of leading logarithms to obtain a nonperturbative
result for the large $r$ potential, 
\begin{equation}
\Phi(r) \longrightarrow -\frac{e}{4\pi r} \times 
\frac1{\sqrt{1 \!-\! \frac{4\alpha}{3\pi} \ln(\frac{R}{r})} } \; . 
\label{RGresum}
\end{equation}

\subsection{Inflationary particle production}

Let us assume that inflation is driven by a minimally coupled scalar 
potential model (\ref{GR+MMCS}). The most transparent gauge is that of 
Salopek, Bond and Bardeen \cite{Salopek:1988qh} in which the temporal condition
sets $\varphi(t,\vec{x}) = \varphi_0(t)$ and the spatial conditions are the
transversality of the graviton. In this gauge the metric components $g_{0\mu}$
are constrained and the dynamical variables $\zeta(t,\vec{x})$ and 
$h_{ij}(t,\vec{x})$ appear in the spatial components,
\begin{equation}
g_{ij}(t,\vec{x}) = a^2(t) e^{2 \zeta(t,\vec{x})} \!\times\! \Bigl[ 
e^{h(t,\vec{x})} \Bigr]_{ij} \qquad , \qquad h_{ii}(t,\vec{x}) = 0 \; . 
\label{gij}
\end{equation}
The quadratic part of the gauge fixed and constrained Lagrangian is,
\begin{equation}
\mathcal{L}^{(2)} = \frac{\epsilon a^3}{8\pi G} \Bigl( \dot{\zeta}^2 -
\frac1{a^2} \vec{\nabla} \zeta \!\cdot\! \vec{\nabla} \zeta \Bigr) +
\frac{a^3}{64 \pi G} \Bigl( \dot{h}_{ij} \dot{h}_{ij} - \frac1{a^2}
h_{ij ,k} h_{ij , k}\Bigr) \; . \label{L2}
\end{equation}

Although the factors of $a(t)$ and $\epsilon(t)$ in (\ref{L2}) break time 
translation invariance, spatial translation invariance is still present. 
This means that the scalar and the graviton fields can be decomposed into 
spatial plane waves. The equation of motion, Wronskian and asymptotic early 
time form for the scalar mode functions $v(t,k)$ are, 
\begin{equation}
\ddot{v} + \Bigl(3 H \!+\! \frac{\dot{\epsilon}}{\epsilon}\Bigr) \dot{v}
+ \frac{k^2 v}{a^2} = 0 \; , \; v \dot{v}^* - \dot{v} v^* = 
\frac{i}{\epsilon a^3} \; , \; v \longrightarrow \frac{\exp[-ik \int_{t_i}^t
\frac{dt'}{a(t')}]}{\sqrt{2 k \epsilon(t) a^2(t)}} . \label{veqns}
\end{equation}
The tensor mode functions $u(t,k)$ obey very similar relations,
\begin{equation}
\ddot{u} + 3 H \dot{u} + \frac{k^2 u}{a^2} = 0 \; , \; u \dot{u}^* - 
\dot{u} u^* = \frac{i}{a^3} \; , \; u \longrightarrow \frac{\exp[-ik \int_{t_i}^t
\frac{dt'}{a(t')}]}{\sqrt{2 k a^2(t)}} . \label{ueqns}
\end{equation}
During inflation $a(t)$ grows nearly exponentially, whereas $H(t)$ is almost
constant. The result is that modes which are originally sub-horizon with
$k < H(t) a(t)$ eventually experience first horizon crossing at $k = H(t_k) a(t_k)$.
One can see from the mode equations (\ref{veqns}-\ref{ueqns}) that the $k^2/a^2$ 
terms become irrelevant after horizon crossing, which causes $v(t,k)$ and $u(t,k)$ 
to approach constants plus deviations falling off like $k^2/a^2$. These constants 
determine the scalar and tensor power spectra,
\begin{equation}
\Delta^2_{\mathcal{R}}(k) = 4\pi G \!\times\! \frac{k^3}{2\pi^2} \!\times\!
\Bigl\vert v(t,k)\Bigr\vert^2_{t \gg t_k} \; , \; \Delta^2_{h}(k) =
32\pi G \!\times\! \frac{k^3}{2\pi^2} \!\times\! 2 \!\times\! \Bigl\vert
u(t,k) \Bigr\vert^2_{t \gg t_k} . \label{power}
\end{equation} 
 
The spatial translation invariance of (\ref{L2}) means that spatially Fourier
transformed fields with wave vector $\vec{k}$ behave as independent harmonic 
oscillators. However, the associated masses and frequencies are time dependent,
\begin{eqnarray}
\zeta & \Longrightarrow & m(t) \sim \epsilon(t) a^3(t) \quad ,\quad 
\omega(t,k) = \frac{k}{a(t)} \; , \label{SHOzeta} \\
h_{ij} & \Longrightarrow & m(t) \sim a^3(t) \qquad ,\quad \omega(t,k) = 
\frac{k}{a(t)} \; . \label{SHOhij}
\end{eqnarray}
The fact that spatial plane waves are independent harmonic oscillators means
that the spectrum of energies at any instant of time is $(N + \frac12) \hbar 
\omega(t,k)$. However, the time dependence of (\ref{SHOzeta}-\ref{SHOhij}) 
means that the energy eigenstates at one time do not remain eigenstates. The
usual ``vacuum'' state is the one which was minimum energy ($N = 0$) in the
distant past. The expectation value of the scalar and tensor energies at
later times take the forms,
\begin{eqnarray}
\Bigl\langle \Omega \Bigl\vert E_{\zeta}(t,k) \Bigr\vert \Omega \Bigr \rangle
& = & \frac12 \epsilon(t) a^3(t) \Bigl[ \vert \dot{v}(t,k) \vert^2 + 
\frac{k^2}{a^2(t)} \vert v(t,k)\vert^2 \Bigr] \; , \qquad \label{Ezeta1} \\
& \longrightarrow & \frac{k}{a(t)} \!\times\! \frac{\pi \Delta^2_{
\mathcal{R}}(k)}{4 G k^2} \!\times\! \epsilon(t) a^2(t) \; , \qquad 
\label{Ezeta2} \\
\Bigl\langle \Omega \Bigl\vert E_{h}(t,k) \Bigr\vert \Omega \Bigr \rangle
& = & \frac12 a^3(t) \Bigl[ \vert \dot{u}(t,k) \vert^2 + \frac{k^2}{a^2(t)} 
\vert u(t,k)\vert^2 \Bigr] \; , \label{Ehij1} \qquad \\
& \longrightarrow & \frac{k}{a(t)} \!\times\! \frac{\pi \Delta^2_{h}(k)}{
64 G k^2} \!\times\! a^2(t) \; . \qquad \label{Ehij2} 
\end{eqnarray} 
If we define occupation numbers based on these energies being $(\frac12 + N) 
\hbar \omega$, the numbers of inflationary scalars and gravitons with a
single super-horizon wave vector $\vec{k}$ are,
\begin{equation}
N_{\zeta}(t,k) = \frac{\pi \Delta^2_{\mathcal{R}}(k)}{4 G k^2} \times 
\epsilon(t) a^2(t) \qquad , \qquad N_{h}(t,k) = 
\frac{\pi \Delta^2_{h}(k)}{32 G k^2} \times a^2(t) \; . \label{numbers}
\end{equation}
Of course there are many wave vectors so the amount of inflationary
particle production is truly staggering.

\subsection{Corrections to EM and GR during inflation}

The best way of understanding quantum loop effects is through the action 
of classical physics on virtual particles. In view of the vast numbers 
(\ref{numbers}) of scalars and gravitons produced out of vacuum, it seems
inevitable that quantum effects are strengthened during inflation. By 
solving the linearized effective field equations we can see how inflationary 
scalars and gravitons modify particle kinematics and long range forces. 

One studies how inflation affects electromagnetism by computing graviton 
and scalar contributions to the vacuum polarization $i[\mbox{}^{\mu} 
\Pi^{\nu}](x;x')$ and then using this to quantum-correct Maxwell's 
equations,
\begin{equation}
\partial_{\nu} \Bigl[ \sqrt{-g} \, g^{\nu\rho} g^{\mu\sigma} F_{\rho\sigma}(x)
\Bigr] + \int \!\! d^4x' \, \Bigl[\mbox{}^{\mu} \Pi^{\nu}\Bigr](x;x') A_{\nu}(x') 
= J^{\mu}(x) \; . \label{EMQFT}
\end{equation}
The one loop graviton contribution to $i[\mbox{}^{\mu} \Pi^{\nu}](x;x')$ was
computed on de Sitter background \cite{Leonard:2013xsa} using the simplest 
gauge \cite{Tsamis:1992xa,Woodard:2004ut}. The result was employed to show 
that the electric fields of plane wave photons experience a secular 
enhancement \cite{Wang:2014tza},
\begin{equation}
F^{\rm 1loop}_{0i}(t,k) \longrightarrow \frac1{\pi} G H^2 \ln(a) \times 
F^{\rm tree}_{0i}(t,k) \; . \label{photon}
\end{equation}
Equation (\ref{EMQFT}) also implies that the response to a point charge
experiences a logarithmic running \cite{Glavan:2013jca},
\begin{equation}
\Phi(t,r) = \frac{Q}{4\pi a r} \Biggl\{ 1 + \frac{G}{3\pi a^2 r^2} + 
\frac1{\pi} G H^2 \ln(a H r) + O(G^2) \Biggr\} \; . \label{Coulomb}
\end{equation}
Note that both of these effects become nonperturbatively strong at late 
times and, in the case of (\ref{Coulomb}), at large $r$.

Quantum corrections to the linearized Einstein equation come from the
graviton self-energy $-i[\mbox{}^{\mu\nu} \Sigma^{\rho\sigma}](x;x')$,
\begin{equation}
\sqrt{-g} \mathcal{L}^{\mu\nu\rho\sigma} h_{\rho\sigma}(x) - \int \!\!
d^4x' \, \Bigl[\mbox{}^{\mu\nu} \Sigma^{\rho\sigma}\Bigr](x;x') 
h_{\rho\sigma}(x') = 8 \pi G T^{\mu\nu}(x) \; , \label{GRQFT}
\end{equation}
where $\mathcal{L}^{\mu\nu\rho\sigma}$ is the Lichnerowicz operator in
the appropriate background. The graviton self-energy was early computed
in the simple gauge \cite{Tsamis:1996qk}. However, this result was not
dimensionally regulated and fully renormalized, so it cannot be used 
in (\ref{GRQFT}). The equation was solved in the Hartree approximation 
to show that the electric curvature components of plane wave gravitons
experience a secular de-enhancement \cite{Mora:2013ypa},
\begin{equation}
C^{\rm 1loop}_{0i0j}(t,k) \longrightarrow -\frac8{\pi} G H^2 \ln(a) \times 
C^{\rm tree}_{0i0j}(t,k) \; . \label{graviton}
\end{equation}
The contribution to $-i[\mbox{}^{\mu\nu} \Sigma^{\rho\sigma}](x;x')$ from
massless, minimally coupled scalars has been computed using dimensional
regularization and fully renormalized \cite{Park:2011ww}. This result 
shows no corrections to dynamical gravitons \cite{Park:2011kg} but it 
does lead to a logarithmic decrease --- in time and space --- in the 
response to a point mass \cite{Park:2015kua},
\begin{equation}
\Psi(t,r) = -\frac{G M}{a r} \Biggl\{ 1 + \frac{G}{20\pi a^2 r^2} - 
\frac{G H^2}{10 \pi} \Bigl[ \frac13 \ln(a) + 3 \ln(a H r)\Bigr] + O(G^2) 
\Biggr\} \; . \label{Newton}
\end{equation}
Like their electromagnetic counterparts, perturbation theory breaks down
for both (\ref{graviton}) and (\ref{Newton}).

\subsection{$\Lambda$-driven inflation}

We have just seen that quanta produced during inflation interact with 
external particles. It is obvious they must also interact with themselves.
Because gravity is attractive, one might expect that they attract one 
another and that this mutual attraction should act as a sort of friction,
slowing down the expansion rate. In line with the factors of $\ln(a)$
seen in expressions (\ref{photon}-\ref{Newton}) one might expect that this 
friction grows with time as more and more of the newly produced quanta
come into causal contact with one another. Also in line with 
(\ref{photon}-\ref{Newton}), we might expect that the secular slowing 
eventually becomes nonperturbatively strong.

No one knows what happens beyond perturbation theory but if we assume that
secular slowing can eventually {\it stop} inflation, it becomes possible
to imagine dispensing with the scalar altogether and driving inflation with 
a large, positive cosmological constant which is gradually screened by the
build up of gravitational self-interaction between inflationary gravitons 
\cite{Tsamis:1996qq,Tsamis:2011ep}. Such a model would solve many of the 
fine tuning problems associated with getting inflation to start, and to 
last long enough, and it would incidentally resolve the old problem of the 
cosmological constant \cite{Weinberg:1988cp,Carroll:2000fy}. It would also 
provide a unique model of inflation which made testable predictions --- if 
only a way could be found to compute in the nonpertubative regime, as we 
did for massless QED in (\ref{RGresum}). 

\section{Answers to My Critics}

These effects (\ref{photon}-\ref{Newton}) are astonishing, and the fact that they 
grow nonperturbatively strong is pregnant with possibilities for late time 
modifications of gravity. None of these results was easy to obtain. Some 
of the computations required the better part of a year's dedicated labor by me 
and/or collaborators. So it is disheartening to watch our work facilely dismissed. 
For example, three of us attended the 12-week KITP program {\it Quantum Gravity 
Foundations: UV to IR} in 2015. Infrared quantum gravitational effects during 
primordial inflation is a subject on which we have a fair claim to being world
experts --- but the world is not interested. Shun-Pei Miao was allotted five 
minutes to summarize her year-long computation showing that inflationary gravitons 
excite fermions by an amount which eventually becomes nonperturbatively large 
\cite{Miao:2005am,Miao:2006gj}. Tomislav Prokopec was given ten minutes to review
his one and {\it two} loop work on scalar quantum electrodynamics during inflation 
\cite{Prokopec:2002uw,Prokopec:2006ue,Prokopec:2008gw}. In this section I will 
discuss four of the reasons which my critics give for their sublime indifference.

\subsection{``{\it Your effects are gauge dependent}''}

First, this objection does not apply to the screening of gravity (\ref{Newton})
caused by loops of massless, minimally coupled scalars \cite{Park:2011ww,Park:2015kua}.
Gauge dependence requires that a graviton propagator enter the loop, and none does
in that case. Yet the effect on the Newtonian potential (\ref{Newton}) shows the 
same fractional correction of $G H^2 \ln(a Hr)$ that gravitons make to the Coulomb
potential (\ref{Coulomb}). I have already invoked Coleridge's famous comment on the
willing suspension of disbelief, which surely applies to dismissing (\ref{Coulomb})
as a gauge artifact while (grudgingly) admitting the reality of (\ref{Newton}). A
second, but closely related point concerns the fractional corrections of $G/a^2 r^2$
which are visible in both (\ref{Coulomb}) and (\ref{Newton}). These are nothing but
the de Sitter descendants, with $r \rightarrow a(t) r$, of flat space corrections 
which were computed long ago \cite{AFR}. 

My third comment concerns the physics of kinematical changes in photons (\ref{photon})
and gravitons (\ref{graviton}). Without regard to the details of the computations,
the vast numbers of particles produced during inflation {\it must} scatter external
photons and gravitons to some extent. This is not some sort of gauge chimera; in flat
space background it is the basis of the proposal to use pulsar arrival times to detect 
gravitational radiation \cite{Detweiler:1979wn}. And the eventual breakdown of 
perturbation theory evident in (\ref{photon}) and (\ref{graviton}) has a very simple
origin: the longer an external photon or graviton propagates the more it should be
affected. The numerical coefficients might be suspect, but the general trend {\it must}
occur.

Of course computational details do matter because we want gauge independent results for
the numerical coefficients. My fourth comment is that years of study have paid off in 
providing both a physical explanation for why the effective field equations are gauge 
dependent and a procedure for eliminating this gauge dependence \cite{Miao:2017feh}.
Gauge dependence arises because some physical source disturbs the effective field, and 
some physical observer measures the disturbance. The source and observer interact with 
quantum gravity, as does everything, and we have not described a physical process unless 
we include this interaction. As might be expected, few of the source and observer
details matter much. For example, it does not matter that the observer had blue eyes
or brown.

Miao, Prokopec (who were granted 15 minutes at a 12-week program!) and I worked 
out an explicit example in flat space background concerning the one graviton loop 
correction to the exchange potential of a massless, minimally coupled scalar in 
the 2-parameter family of covariant gauge fixing functions \cite{Miao:2017feh},
\begin{equation}
\mathcal{L}_{GF} = -\frac1{2 a} \eta^{\mu\nu} F_{\mu} F_{\nu} \qquad , \qquad
F_{\mu} = \eta^{\rho\sigma} \Bigl( h_{\mu \rho , \sigma} - \frac{b}{2} 
h_{\rho\sigma , \mu} \Bigr) \; . \label{gauge}
\end{equation}
The linearized effective field equation can be expressed in terms of a self-mass-squared
$M^2(x;x')$,
\begin{equation}
\partial^2 \varphi(x) - \int \!\! d^4x' \, M^2(x;x') \varphi(x') = J(x) \; . \label{EFE}
\end{equation}
Ignoring the source and observer, our result for $M^2(x;x')$ in gauge (\ref{gauge}) 
takes the form of a gauge dependent constant times a function of spacetime whose form
is fixed by Poincar\'e invariance and dimensionality,
\begin{eqnarray}
-i M^2(x;x') & = & \mathcal{C}_0(a,b) \times \frac{G \partial^6}{4 \pi^3} \Biggl[ 
\frac{ \ln(\mu^2 \Delta x^2)}{\Delta x^2} \Biggr] \quad , \quad \Delta x^2 \equiv 
(x \!-\! x')^2 \; , \label{M0} \qquad \\
\mathcal{C}_0(a,b) & = & +\frac34 -\frac34 \times a - \frac32 \times \frac1{b \!-\! 2}
+ \frac34 \times \frac{(a \!-\! 3)}{(b \!-\! 2)^2} \; . \label{C0}
\end{eqnarray}

The fact that $\mathcal{C}_0(a,b)$ can be made to vary from $-\infty$ to $+\infty$ 
would provide my critics justification to condemn the whole exercise as nonsense. 
However, they would be wrong! By including quantum gravitational correlations from 
the source and observer one gets additional contributions having the same spacetime 
dependence as (\ref{M0}) but with different gauge dependent coefficients.
Table~\ref{Cab} summarizes the results and demonstrates the satisfying cancellation 
of all dependence on the parameters $a$ and $b$ \cite{Miao:2017feh}. Steven Weinberg 
relates how post-renormalization physicists used to quip, ``just because something 
diverges doesn't mean it's zero.'' In the same vein, I hope people will now admit, 
``just because something depends on the gauge doesn't mean it's zero!'' 
\begin{table}[H]
\setlength{\tabcolsep}{8pt}
\def\arraystretch{1.5}
\centering
\begin{tabular}{|@{\hskip 1mm }c@{\hskip 1mm }||c|c|c|c|c|}
\hline
$i$ & $1$ & $a$ & $\frac1{b-2}$ & $\frac{(a-3)}{(b-2)^2}$ &
{\rm Description} \\
\hline\hline
0 & $+\frac34$ & $-\frac34$ & $-\frac32$ & $+\frac34$ &
{\rm scalar\ exchange} \\
\hline
1 & $0$ & $0$ & $0$ & $+1$ & {\rm vertex-vertex} \\
\hline
2 & $0$ & $0$ & $0$ & $0$ & {\rm vertex-source,observer} \\
\hline
3 & $0$ & $0$ & $+3$ & $-2$ & {\rm vertex-scalar} \\
\hline
4 & $+\frac{17}4$ & $-\frac34$ & $0$ & $-\frac14$ &
{\rm source-observer} \\
\hline
5 & $-2$ & $+\frac32$ & $-\frac32$ & $+\frac12$ &
{\rm scalar-source,observer} \\
\hline\hline
Total & $+3$ & $0$ & $0$ & $0$ & \\
\hline
\end{tabular}
\caption{The gauge dependent factors $C_i(a,b)$ for each contribution to
the self-mass-squared.}
\label{Cab}
\end{table}

Resolving the gauge problem for effective field equations is not enough.
The correct generalization of the power spectrum is still unclear 
\cite{Miao:2013oko} but the answer may be gauge invariant correlators.
There is interesting recent work on these by Markus Fr\"ob and
collaborators \cite{Frob:2017lnt,Frob:2017gyj,Frob:2018lfo}. An invariant 
measure of the local expansion rate has also been proposed 
\cite{Tsamis:2013cka}, and its one loop renormalization on de Sitter 
background has been accomplished \cite{Miao:2017vly}.

\subsection{``{\it IR gravitons have small curvature}''}

Each graviton mode has a constant wave vector $\vec{k}$, whose physical 
wave vector $k/a(t)$ redshifts as the universe expands. During a sufficiently
long epoch of inflation the physical wave number eventually falls below the
almost constant Hubble parameter $H(t)$, an event which is known as {\it first
horizon crossing}. After inflation the product $a(t) H(t)$ falls off, so modes
can experience a {\it second horizon crossing}. Some of my critics appear to 
believe that only sub-horizon gravitons are physical, so that inflationary 
gravitons literally pass out of existence between first and second crossings.
This is nonsense. The curvatures of super-horizon gravitons become small 
exponentially fast, but they never touch zero, and the exponentially growing 
numbers of super-horizon gravitons (\ref{numbers}) can compensate small 
individual curvatures. 

Another point of great relevance is that the curvatures of inflationary gravtions
are not always small; they start out large and then redshift. General relativity 
on asymptotically flat space has a well-known {\it nonlinear memory effect} 
\cite{Christodoulou:1991cr} in which the passage of a gravitational wave leaves 
a permanent displacement in test observers. After the wave has passed, the 
observers' curvature is {\it zero}, yet the passage of the wave had an effect. 
So how can it be argued that the continual redshift of gravitons from the 
ultraviolet to the infrared during inflation can have no effect?

I seem to be re-contesting the same battles which were fought last
century over the fact that charged quantum particles couple to the undifferentiated
vector potential, not the field strength. There never was any doubt about this,
but stubborn physicists for years resisted the obvious conclusion that constant
vector potentials could, under certain circumstances, engender physical effects.
But nature pays no attention to human prejudices and an experiment was eventually 
proposed \cite{ES,Aharonov:1959fk} whose result \cite{RGC} is taught in undergraduate 
quantum mechanics. In the same sense matter, and gravity itself, couple to the 
undifferentiated metric, not to the curvature. Hence there must be cases in which
physical effects can occur even for zero curvature.

Finally, one must distinguish between infrared {\it divergences} and infrared 
{\it effects} such as secular growth (\ref{photon}) and logarithmic running 
(\ref{Coulomb}). Infrared divergences derive from infinite numbers of gravitons 
being super-horizon at the start of inflation. A compelling case has been made 
that these would be subsumed into the background of any local observer and have 
no effect \cite{Giddings:2011zd}. On the other hand, infrared effects are caused 
by gravitons which were initially sub-horizon and experienced first horizon
crossing during inflation. It is not legitimate to regard these are having
always been part of the background. Of course it might still have been correct,
as my critics insisted. I'm a big believer in checking things when I can figure 
out how to do it. When I finally figured out how to check this belief, the 
result is that is that subsuming infrared gravitons into the background can
neither eliminate secular growth  \cite{Basu:2016iua} nor changes in the local\
expansion rate \cite{Basu:2016gyg}.

\subsection{``{\it Your effects are not observable}''}

Some claim that the cosmological horizon precludes local observers from perceiving 
effects like (\ref{photon}-\ref{Newton}) {\it during} inflation. To see how a  
local observer could perceive the secular enhancement of photons (\ref{photon}) 
consider setting off a flash between reflecting mirrors at fixed, sub-horizon 
physical distances from one another, and monitoring the field strength as the 
signal reflects back and forth. The logarithmic running of (\ref{Coulomb}) could 
be observed using one neutral and two charged particles. Release the two charges 
at rest from one another, with the neutral particle also released from rest, next 
to one of the charges. Then even after the two charges are no longer in causal 
contact, the effect of their mutual attraction could be followed by measuring the 
separation of the nearby charge from the neutral particle. These sorts of 
experiments could at least be done for a while.

People who make the second objection leave me wondering what part of the phrase
``nonperturbatively strong'' they do not understand. The burden of my message 
is that modifications to gravity {\it now} might derive from interactions 
between the vast numbers (\ref{numbers}) of scalars and gravitons produced 
during inflation. The effects (\ref{photon}-\ref{Newton}) all grow to eventually 
become nonperturbatively strong during a prolonged period of inflation. The fact 
that I cannot yet sum up the series of large logarithms to exhibit their late 
time limits is no reason for claiming that those limits are unobservable.

\subsection{``{\it Your calculations are difficult}''}

No one enjoys being attacked, but I have tried very hard to understand
my critics and to honestly address their concerns. In a well-known case 
I publicly renounced a previous opinion \cite{Abramo:1998hi} based on
this sort of engagement \cite{Abramo:2001dc}. However, whining about the
complexity of perturbative quantum gravity on de Sitter background is not 
a serious objection, although it is sadly frequent, and I can only urge 
this class of critics to {\bf grow up!}

\section{Conclusions}

This article has been devoted to making the case for nonlocal modifications 
of gravity. In these concluding remarks I will briefly review some of the
models I have explored. I apologize to the many others whose work on such
nonlocal models I will not discuss \cite{Parker:1985kc,Banks:1988je,
Wetterich:1997bz,Barvinsky:2003kg,Espriu:2005qn,Hamber:2005dw,Biswas:2005qr,
LopezNacir:2006tn,Khoury:2006fg,Capozziello:2008gu,Biswas:2010zk,Zhang:2011uv,
Barvinsky:2011hd,Barvinsky:2011rk,Elizalde:2011su,Barvinsky:2012ts,
Biswas:2013cha,Foffa:2013sma,Foffa:2013vma,Kehagias:2014sda,Maggiore:2014sia,
Dirian:2014ara,Conroy:2014eja,Barreira:2014kra,Dirian:2014xoa,Dirian:2014bma,
Zhang:2015ahe,Netto:2015cba,Cusin:2015rex,Zhang:2016ykx,Cusin:2016nzi,
Dirian:2016puz,Maggiore:2016fbn,Koshelev:2016xqb,Nersisyan:2016hjh,
Maggiore:2016gpx,Wu:2016evd}.

As I have explained, nonlocal modifications of gravity can come from quantum
loop effects that grew nonperturbatively strong during primordial inflation. 
Although one can follow some of these effects while they are still small,
there is not yet any way to compute past the breakdown of perturbation theory
as we were able to do for the Coulomb potential (\ref{RGresum}) of massless 
QED in flat space. However, the putative inflationary origin does explain two
things that would otherwise seem unnatural:
\begin{itemize}
\item{There is an initial value surface upon which the initial conditions
of inverse differential operators can be defined; and}
\item{Modifications of gravity are expected on large distances, not small ones.}
\end{itemize}

We seek to guess the most cosmologically significant part of the gravitational
effective action. What happens perturbatively should serve as a guide. One can
see from expressions (\ref{photon}-\ref{Newton}) that secular growth on de 
Sitter background resides in factors of $\ln[a(t)]$. A simple nonlocal scalar
which reproduces this is \cite{Romania:2012av},
\begin{equation}
\frac1{\square} R\Bigr\vert_{\rm dS} \longrightarrow -4 \ln[a(t)] \; ,
\label{guess}
\end{equation}
where $\square \equiv \frac1{\sqrt{-g}} \partial_{\mu} (\sqrt{-g} \, g^{\mu\nu}
\partial_{\nu})$ is the scalar d`Alembertian. So it might be reasonable to
expect that modifications involve an algebraic function $f(\frac1{\square} R)$.
Models of this type have been proposed to study $\Lambda$-driven inflation
\cite{Tsamis:1997rk,Tsamis:2009ja}, metric realizations of MOND 
\cite{Soussa:2003vv} and late time acceleration \cite{Deser:2007jk,
Deser:2013uya,Woodard:2014iga}. Although this simple ansatz is not satisfactory
for MOND \cite{Soussa:2003sc} it does describe an interesting model for ending
primordial inflation, generating density perturbations and then reheating to
go quiescent into the epoch of radiation domination \cite{Tsamis:2010pt}. 
With some elaborations it might even describe late time acceleration
\cite{Tsamis:2010ph,Tsamis:2014hra,Tsamis:2016boj}. The simple ansatz also 
offer many advantages in describing late time acceleration because:
\begin{itemize}
\item{Unlike $R \longrightarrow f(R)$ models, theories involving $R \longrightarrow
R f(\frac1{\square} R)$ can be chosen to exactly reproduce the $\Lambda$CDM
expansion history \cite{Deffayet:2009ca};}
\item{Because $\frac1{\square} R$ is {\it negative} for cosmology and {\it 
positive} for gravitationally bound systems, it is trivial to choose the
function $f(\frac1{\square} R)$ to avoid solar system constraints;}
\item{The scalar $\frac1{\square} R$ is dimensionless so it requires no small
mass;}
\item{During radiation domination $R = 0$, so the onset of modifications
is postponed until late in cosmological history;}
\item{Even after matter domination, the scalar $\frac1{\square} R$ only
grows logarithmically with time, postponing the onset to even later times;
and}
\item{Perturbing the model to study structure formation \cite{Park:2012cp,
Dodelson:2013sma,Park:2016jym} actually agrees better with the data than 
general relativity \cite{Nersisyan:2017mgj,Park:2017zls}.}
\end{itemize}

A more elaborate nonlocal model involving an algebraic function of a 
different scalar has been devised to reproduce MOND pheneomenology for 
gravitationally bound systems \cite{Deffayet:2011sk}. Because this is a 
complete, metric theory of gravity it can be applied to cosmology just
like general relativity \cite{Deffayet:2014lba,Woodard:2014wia}. The
algebraic function can be chosen to reproduce most of the $\Lambda$CDM 
expansion history, and even offers a serendipitous explanation for the
tension between low redshift and high redshift determinations of the
Hubble constant \cite{Kim:2016nnd}. However, perturbations about the 
cosmological background do not correctly describe structure formation
\cite{Tan:2018bfp}. I suspect that the problem can be resolved by making
the numerical coincidence $c H_0 \simeq 2 \pi a_0$ dynamical.

\centerline{\bf Acknowledgements}

I am grateful for conversations and correspondence with S. Deser, E.
Kiritsis and N. C. Tsamis. This work was partially supported by NSF 
grants PHY-1506513 and PHY-1806218, and by the Institute for 
Fundamental Theory at the University of Florida.

\end{document}